\documentclass[double]{article}
\usepackage{times}
\usepackage{algorithm}
\usepackage{algorithmic}
\usepackage{wrapfig}
\usepackage{verbatim}
\usepackage{amsmath}
\usepackage{graphicx}
\usepackage{subcaption}
\usepackage[active]{srcltx}
\usepackage{color}
\usepackage{lineno}
\usepackage{dirtytalk}
\usepackage{amsthm}
\usepackage{authblk}
\usepackage{listings}
\usepackage{tikz}
\usetikzlibrary{shapes.geometric, arrows}

\usepackage{epsfig,graphicx,amsfonts}
\usepackage{amsmath,amsthm,amssymb}
\usepackage{xcolor}

\usepackage{adjustbox}
\usepackage{multirow}
\usepackage{setspace}
\usepackage{hyperref}
\usepackage{comment}

\long\def\comment #1\commentend{}

\begin{document}

\title{\Large The Scientometrics and Reciprocality Underlying Co-Authorship Panels in Google Scholar Profiles}

\author{Ariel Alexi$^{1*}$, Teddy Lazebnik$^{2x}$ and Ariel Rosenfeld$^{1x}$\\
\(^1\) Department of Information Science, Bar Ilan University, Ramat Gan, Israel\\
\(^2\) Department of Cancer Biology, Cancer Institute, University College London, London, UK\\
\(^*\) Corresponding author: ariel.147@gmail.com \\ 
\(^x\) These authors contributed equally 

}

\maketitle 

\date{ }

\begin{abstract}
Online academic profiles are used by scholars to reflect a desired image to their online audience. In Google Scholar,  scholars can select a subset of co-authors for presentation in a central location on their profile using a social feature called the \say{Co-authroship panel}. In this work, we examine whether scientometrics and reciprocality can explain the observed selections. To this end, we scrape and thoroughly analyze a novel set of 120,000 Google Scholar profiles, ranging across four disciplines and various academic institutions. Our results suggest that scholars tend to favor co-authors with higher scientometrics over others for inclusion in their co-authorship panels. Interestingly, as one’s own scientometrics
are higher, the tendency to include co-authors with high scientometrics is diminishing. Furthermore, we find that reciprocality is central in explaining scholars' selections.
\end{abstract}

\maketitle \thispagestyle{empty}

\pagestyle{myheadings} \markboth{Draft:  \today}{Draft:  \today}
\setcounter{page}{1}

\section{Introduction}
\label{sec:introduction}
In today's digital age, an individual's online presence plays a central role in shaping one's identity and reputation within society \cite{intro_1,intro_2}. The increasing popularity of social media platforms is a testament to this phenomenon, as people strive to project a desired image to their online audience \cite{ashraf2023antecedents, petroni2019social}. Within the academic community, online academic profiles have become essential reflections of scholars' professional identities \cite{intro_3,intro_4}. Unfortunately, many academic profiles, and especially those (semi-)automatically generated by various bibliometric services, offer limited freedom for scholars in terms of self-information disclosure, leading scholars to carefully select which information to present \cite{intro_5}. One intriguing aspect of this self-information disclosure, which to the best of our knowledge has yet to be examined in the literature, is the selection of co-authors one wishes to highlight in his/her academic profile. 

In the realm of social psychology, self-identification theory \cite{goffman2016presentation, baumeister1982self} provides a theoretical framework for understanding how individuals shape their public image and manage their self-presentation in various social contexts \cite{self_presentation_fashion,self_presentation_management_orga,self_presentation_athletes}. While several variants of the theory have been posed over the years, most proposals seem to agree on several fundamental concepts including the notion that individuals strive to present themselves in a way that aligns with their desired self-image, taking into account the perceived expectations and norms of the audience they wish to impress or influence \cite{self_presentation_1, self_presentation_2,self_presentation_3, self_presentation_4,self_presentation_5, self_presentation_6,self_presentation_7}.
In the academic context, scholars' online profiles serve as a crucial platform for self-presentation, offering scholars the opportunity to curate an image they wish to project to their peers, potential collaborators, and even funding bodies \cite{self_presentation_academia, self_presentation_students_academic_achievement}. Thus, conceivably, scholars may wish to enhance their own perceived image by aligning themselves with strong and prominent co-authors. In many ways, this expectation to benefit from one's association with others is also connected to the sociological concept of \say{social capital} \cite{rosa_1}. That is, social connections and interactions are assumed to have inherent value and can thus lead to various positive outcomes in personal and professional settings alike. Indeed, prior work has established that collaboration with leading researchers can positively impact one's academic career \cite{amjad2017standing, coauthor_prestige_geoloc_displine, author_prestige_effect_your_work}. 
However, the choice to \textit{highlight} specific co-authors in one's academic profile, as opposed to the \textit{actual collaboration} with that co-author, need not necessarily follow the same pattern. First, various social and professional norms such as \textit{reciprocally} may also govern this unique socio-academic dynamic. That is, when one scholar includes another in his/her co-authorship panel, the other may feel pressured to reciprocate and do the same, as is often encountered in various professional and social interactions \cite{friends_affect_political,friends_affect_political_2}. 
In addition, the value derived from the collaboration may be traced back to the learning and mentoring experienced through that collaboration and significantly less to the association itself presented in one's profile \cite{rosa_2}. 

It is important to note that scholars may have different levels of control over the information displayed on their academic profile depending on the platform in question. This includes the presentation of publications, achievements, expertise, and collaborative networks, to name a few \cite{ariel_2}. Focusing on Google Scholar\footnote{\url{https://scholar.google.com/}} (GS), arguably the most widely used academic search engine and bibliographic database to date, scholars are provided with a simple interface to manage their academic profile \cite{gusenbauer2019google}. The profile includes a comprehensive list of that scholar's publications, scientometrics, and various meta-data details such as name, field of study, and affiliation \cite{AboutGoogleScohlar}. In addition, the profile includes a social feature that is almost entirely under the scholar's control -- the \textit{co-authorship panel}. This co-authorship panel allows scholars to select a subset of their co-authors, as determined by their indexed publications, to be showcased in a central location on their profile. 

In this work, we examine whether the co-authors selected for inclusion in one's co-authorship panel on GS can be explained by their associated scientometrics (i.e., citation counts, i10-index, and h-index). In addition, we examine the prevalence of reciprocally within this socio-academic dynamic. Methodologically, we scrape and analyze a novel and comprehensive set of roughly 120,000 GS profiles. For each profile, in addition to the provided meta-data of the scholar in question, the entire set of co-authors was extracted and the co-authorship panel was identified.  
Overall, our results seem to suggest that, indeed, scholars tend to favor co-authors with higher scientometrics over others for inclusion in their co-authorship panels. Interestingly, as one's  
own scientometrics are higher, the tendency to include co-authors with high scientometrics is diminishing. Furthermore, we find that reciprocality is central in explaining scholars' panel selections.  

The article is organized as follows: In Section \ref{sec:methods}, we describe our data collection procedures and the adopted analytical approach. Next, in Section \ref{sec:results}, we present our main findings. Following, in Section \ref{sec:discussion}, we discuss our results, interpret them in context, and provide directions for future work.

\section{Materials and Methods}
\label{sec:methods}

\subsection{Data}
In order to obtain a diverse and comprehensive set of scholars, we adopted 359 GS profiles which were retrieved and analyzed by \cite{hodayah_paper} (albeit for other purposes). Per the authors' study design, these profiles cover both high, middle, and low-ranked US universities\footnote{According to the Shanghai Ranking -- \url{https://www.shanghairanking.com/}}, and are associated with four distinct disciplines: Biology (103), Computer Science (101), Philosophy (52), and Psychology (103). 
For each of these 359 scholars, which are at the focus of our subsequent analysis and can be conceptually thought of as \say{seed} scholars, we retrieved and scraped the GS profiles of \textit{all} their co-authors, and all their co-authors' co-authors -- resulting in roughly 120,000 profiles \textit{that had a non-empty co-authorship panel}. In other words, for each scholar, we retrieved the GS profiles of any scholar who either co-authored a publication with that scholar or has co-authored a publication with one of his/her co-authors. Note that all seed scholars had a non-empty co-authorship panel. In addition, any scholar encountered during the retrival procedure who had an empty co-authorship panel was omitted from further consideration (approximately 25.21\% of the encountered profiles). 

For each profile in the resulting dataset (both seed profiles and otherwise), we collected the six scientometrics provided in the GS profile: citation count, i10-index, and h-index; once since 2018 (i.e., last 5 years), and once in total as of June 2023 (i.e., lifetime).
For completeness, we further determine the scholar's academic age and estimate his/her gender. For the academic age, we consider the timespan between one's earliest indexed publication and the current year (2023). As for gender, we adopted the widely used model of \cite{gender_use} which predicts gender according to one's name. In order to avoid ambiguity, we determine one's gender only if the model's confidence is higher than 95\%. The remaining 21.45\% profiles were disregarded.

\subsection{Analytical Approach}

Our analytical approach consists of four phases:
First, we start our investigation with a basic statistical analysis examining whether the scholars included in one's co-authorship panel favorably compare, in scientometric terms, to those who were excluded from it. To that end, for each seed scholar, we compare the mean scientometrics of the co-authors included in the panel to that of the co-authors excluded from it using standard t-tests. In addition, we performed paired t-tests at the population level with each scholar representing a single sample consisting of the mean scientometrics of those included in his/her panel on the one hand and that of those excluded from it on the other. 
Second, we examine the possible connection between one's own academic success metric and his/her alignment with the studied phenomenon. Specifically, we study whether one's tendency to select higher-profile co-authors for inclusion in his/her co-authorship panel is regulated by their own academic success. To that end, for each seed scholar, we compute the ratio between the mean of each scientometric of those included in the panel and that of the entire set of co-authors and examine these ratios' possible relation with the scholar's own scientometrics. Clearly, the higher the ratio in any of the examined scientometrics -- the more often one chooses higher-performing co-authors over others for inclusion in his/her co-authorship panel. These ratios are, in turn, used as dependent variables in separate regression models with the scholar's own scientometric acting as the independent variable and the ratio in question acting as the dependent variable. In order to determine the best fit, we use the SciMed model \cite{scimed}, a symbolic regression tool that searches a large and versatile set of analytical functions given an optimization objective and a dataset. 
Third, we turn to examine the prevalence of reciprocality. We start by examining the prevalence of reciprocal inclusion in the entire population and look for possible differences across disciplines using a chi-square test with post-hoc chi-square tests with Bonferroni corrections. Then, using a machine learning-based analysis, we train and evaluate a Random Forest (RF) model \cite{intro_ml_2} with grid search for the hyperparameter tuning \cite{grid_search}. That is, we examine if, and to what extent, the scientometrics of two scholars who have published together in the past could explain the reciprocal inclusion in each other's co-authorship panels or the lack thereof. The models are trained using a standard 80-20 split to a train (80\%) and test (20\%) cohorts of the dataset \cite{data_split_1,data_split_2} and evaluated using the standard metrics of accuracy, recall, precision, and \(F_1\) score (see \cite{rosa_3} for the relevant statistical background). In order to determine feature importance, we adopted the standard technique used for RF models by removing each feature, one feature at a time, and computing the change in the model's accuracy to obtain the relative importance \cite{rf_importance}. 

\section{Results}
\label{sec:results}

We start by reporting the main statistical characteristics of our dataset. For the entire set of seed scholars (i.e., the sample), the mean academic age is \(19.98\pm 9.98\) with a range between 4 and 39. Considering gender, our data is characterized by approximately 3:1 male-to-female ratio. Considering scientometrics, the mean total citation count is \(11375.36\pm 20608.60\) with a range of \(16-170195\); the mean total h-index is \(36.25\pm 27.18\) with a range of  \(2-158\); and mean total i10-index is \(75.80\pm 94.70\) with a range of \(1-921\). Similarly, for last 5 years' version of the scientometrics, mean citation count is \(4453.63\pm 7960.33\) with a range of \(10-93743\); mean h-index is \(25.13\pm 16.54\) with a range of \(2-139\); and i10-index is \(52.28\pm 55.16\) with a range of \(0-379\).

Next, we consider each seed scholar individually and examine whether the co-authors included in his/her co-authorship panel are statistically associated with  better scientometrics than those excluded from it. As can be seen from Table \ref{tab:hypo_1}, most scholars seem to align with this expectation with higher rates being recorded for the scientometrics calculated throughout the co-authors' lifetime (i.e., total) as opposed to the last 5 years. At the population level, as can be seen in Table \ref{tab:hypo_2}, the results seem to agree with those reported above, with the scientometrics calculated throughout one's lifetime demonstrating high levels of statistical significance, all at \(p < 0.001 \), compared to the those calculated based on the last 5 years which present lower levels of statistical significance, if any.  

\begin{table}[!ht]
    \centering
    \begin{tabular}{lcc}
    \hline \hline 
    \textbf{Metric} & Total \  & Last 5 years \ \\ \hline \hline 
    Citation & \(83\%\) & \(59\%\) \\
    H-index & \(89\%\) & \(66\%\) \\
    i10-index & \(74\%\) & \(52\%\) \\ \hline \hline 
    \end{tabular}
    \caption{The percent of scholars for which the co-authors included in the co-authorship panel significantly outperform, in scientometric terms, those who were excluded from it at $p\leq0.05$. }
    \label{tab:hypo_1}
\end{table}

\begin{table}[!ht]
    \centering
    \begin{tabular}{lcc}
    \hline \hline 
    \textbf{Metric} & Total & Last 5 years \\ \hline \hline 
    Citation & \(1.34 \cdot 10^{-4***}\) & \(0.047^{*}\) \\
    H-index & \(3.53 \cdot 10^{-5***}\) & \(0.009^{**}\) \\
    i10-index & \(6.74 \cdot 10^{-6***}\) & \(0.059\) \\ \hline \hline 
    \end{tabular}
    \caption{The \(p\)-value of a t-test comparing the scientometric in question between the co-authors included in one's co-authorship panel and those who were excluded from it. \(^* \text{ denotes } p \leq 0.05, ^{**} \text{ denotes } p \leq 0.01\text{ and } ^{***} \text{ denotes } p \leq 0.001\)}
    \label{tab:hypo_2}
\end{table}

\begin{figure}[!ht]
    \centering
    \includegraphics[width=0.99\textwidth]{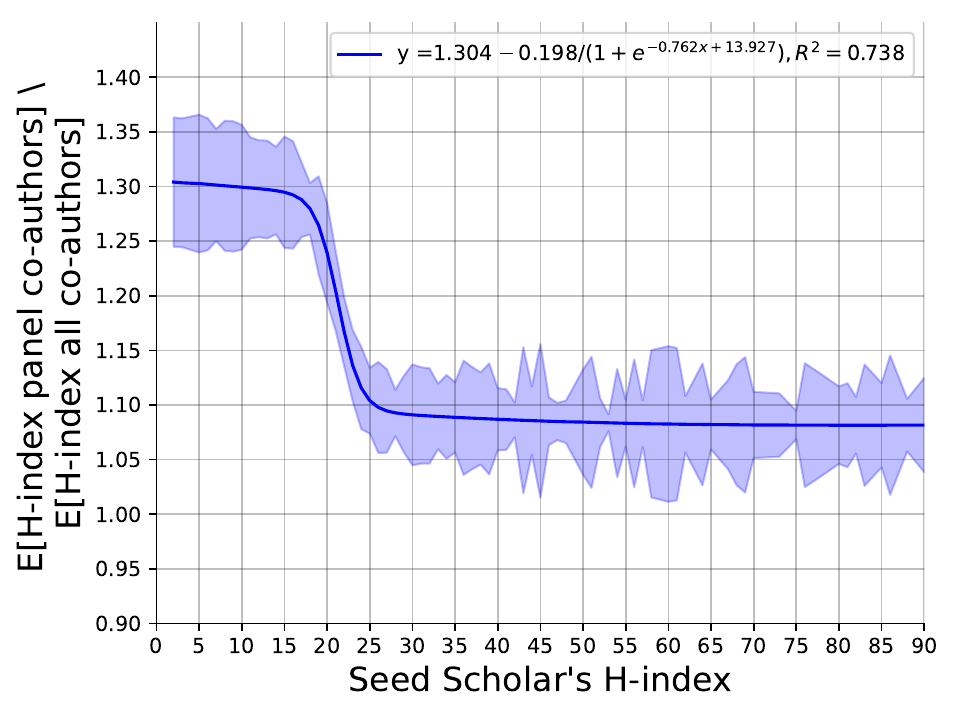}
    \caption{The relation between one's lifetime H-index and the ratio between his/her co-authorship panel's mean life H-index and all his/her co-authors' lifetime H-index. The shaded area denotes the confidence interval which includes 85\% of the data.}
    \label{fig:hypo_3}
\end{figure}

Focusing on individual seed scholars, we consider the possible connection between one's own scientometrics and the ratio between the scientometrics of those included in his/her co-authorship panel and all his/her co-authors. Considering the lifetime H-index as a representative example, 
Fig. \ref{fig:hypo_3} presents a reversed sigmoid relation that seems to emerge from the data with following the mathematical form \(y = 1.304 - \frac{0.198}{1 + e^{-0.762x + 13.927}}\) where \(y\) is the mean lifetime H-index of the panel co-authors divided by the mean lifetime H-index of all the co-authors and \(x\) is the seed scholar's lifetime H-index. The coefficient of determination of this model is considered high, with \(R^2 = 0.738\). Similar patterns, albeit with lower \(R^2\) scores, were identified for the lifetime citations and i10-index (\(R^2\) of 0.417 and 0.386, respectively). Lower \(R^2\) scores were recorded for the last 5 years' citations, H-index and i10-index (\(R^2\) scores of \(0.358\), \(0.512\), and \(0.317\), respectively. 
Similar analyses considering gender, discipline, or academic age, did not yield any statistically significant differences. 

Following, we focus on reciprocality in co-authorship panel inclusion. As shown in Table \ref{tab:mutual_agreement}, a high average of 57.8\% of all inclusions in one's co-authorship panel are reciprocal, i.e., both scholars include each other in their respective panels. When breaking scholars by disciplines, we see that the disciplines differ significantly at $p<0.05$. Specifically, we find that scholars from Biology present considerably higher levels of reciprocity (almost every seven out of ten inclusions -- 69.82\%) compared to all other disciplines. Similarly, scholars from Psychology show significantly higher levels of reciprocity (more than every six out of ten inclusions -- 62.02\%) compared to philosophy (less than half  -- 47.43\%). All other differences are not found to be statistically significant at $p<-0.05$. Similar analyses considering one's own scientometrics, gender, or academic age, did not yield any statistically significant differences. 

\begin{table}[!ht]
\centering
\begin{tabular}{c c c c} 
\hline \hline
 \textbf{Discipline} & \textbf{One-Way} & \textbf{Reciprocal} \\
\hline \hline
 Philosophy       & 52.57\% (175) &   47.43\% (194)\\
 Biology          & 30.18\% (1659) &  69.82\% (717)\\
 Computer Science & 48.08\% (1767) &  51.92\% (1636)\\
 Psychology       & 37.98\% (1703) & 62.02\%  (1043)\\ \hline
 p-value & & 0.017 \\ \hline
 \textbf{Average}       & 42.20\% & 57.80\% \\
    \hline \hline
\end{tabular}
    \caption{Inclusion reciprocality across academic disciplines.}
    \label{tab:mutual_agreement}
\end{table}

Finally, we train an RF model to predict if inclusion in one's co-authorship panel is reciprocal or not using the entire set of features at our disposal (see the first column of Table \ref{tab:classifier_feature_importance}). The results, as shown in Table \ref{tab:classifier_results}, suggest that the trained model is well-suited for the task of predicting if inclusion is reciprocal or not at very high levels of accuracy, recall, precision, and $\boldsymbol{F_1}$ scores (train data: 0.93, 0.88, 0.97 and 0.92; test set: 0.85, 0.8, 0.87 and 0.83, respectively). It is important to note that the model demonstrates only very minor overfitting on the train set compared to the test set. When focusing on feature importance, as shown in the second column of Table \ref{tab:classifier_feature_importance}, the number of mutual inclusions the other scholar has clearly stood out from the rest. In other words, the tendency of other scholars to practice mutual inclusions \textit{regardless of the scholar in question} is very indicative of his/her likelihood to include that scholar as well. When considering feature importance \textit{without this feature}, as shown in the third column of Table \ref{tab:classifier_feature_importance}, we see the scholar's tendency to practice mutual inclusion stands out. In other words, the tendency of a scholar to practice mutual inclusions is very indicative of the likelihood \textit{others} will include him/her as well. Taken jointly, it seems that reciprocal practices, at either or both sides of the interaction, are prominent within the model for determining reciprocality for unseen instances. Interestingly, one's gender, academic age, and own scientometrics seem to play relatively small roles in the model's predictions.  However, aligned with prior results, the \textit{other scholar's scientometrics} do seem to play a larger role within the model as those with higher scientometrics present less reciprocality than others. 

\begin{table}[!ht]
     \centering
\begin{tabular}{llcccc}
\hline \hline
\textbf{Model} & \textbf{Cohort} & \textbf{Accuracy} & \textbf{Recall} & \textbf{Precision} & \(\boldsymbol{F_1}\) \textbf{score} \\ 
 \hline \hline
\multirow{2}{*}{Random Forest} &  Train &  0.93 & 0.88  & 0.97 & 0.92\\
 &   Test  &  0.85 & 0.80 & 0.87 & 0.83 \\
 \hline \hline
\end{tabular}
\caption{The performance of a Random Forest classifier model that predicts if inclusion in one's co-author panel is reciprocal or not.}
\label{tab:classifier_results}
 \end{table}

 \begin{table}[!ht]
     \centering
     \begin{tabular}{lcc}
\hline \hline
 \textbf{Feature}& \textbf{Feature importance} & \textbf{Feature importance (reduced)} \\
\hline \hline
 \(1_{st}\) scholar's gender                             & 0.002 & 0.005 \\ 
 \(1_{st}\) scholar's one way authors count             & 0.037 & 0.099 \\ 
 \(1_{st}\) scholar's mutual inclusion count             & 0.040 & 0.107\\ 
 \(1_{st}\) scholar's number of articles                    & 0.016 & 0.040\\ 
 \(1_{st}\) scholar's number of co-authors & 0.031 & 0.042 \\ 
 \(1_{st}\) scholar's total citations                     & 0.015 & 0.030 \\ 
 \(1_{st}\) scholar's citations from the last 5-years & 0.012 & 0.031 \\ 
 \(1_{st}\) scholar's H-index & 0.013 & 0.029\\ 
 \(1_{st}\) scholar's H-index from the last 5-years               & 0.013 & 0.025\\ 
 \(1_{st}\) scholar's i10-index & 0.014 & 0.032\\ 
 \(1_{st}\) scholar's i10-index from the last 5-years & 0.012 & 0.026 \\ 
 \(1_{st}\) scholar's academic age & 0.010 & 0.023\\ 
 \(2_{nd}\) scholar's gender                   & 0.004 & 0.027 \\ 
 \(2_{nd}\) scholar's one way authors count & 0.033 & 0.015 \\ 
 \(\boldsymbol{2_{nd}}\) \textbf{scholar's mutual inclusion count} & 0.567 & --- \\ 
 \(2_{nd}\) scholar's number of articles  & 0.000 & 0.066\\ 
 \(2_{nd}\) scholar's total citations        & 0.026 & 0.055 \\ 
 \(2_{nd}\) scholar's  citations from the last 5-years & 0.026 & 0.058 \\ 
 \(2_{nd}\) scholar's H-index  & 0.024 & 0.045 \\ 
 \(2_{nd}\) scholar's H-index from the last 5-years & 0.020 & 0.045 \\ 
 \(2_{nd}\) scholar's i10-index & 0.029 & 0.062 \\ 
 \(2_{nd}\) scholar's i10-index from the last 5-years & 0.027 & 0.048 \\ 
 \(2_{nd}\) scholar's academic age  & 0.024 & 0.071 \\ 
\hline \hline
\end{tabular}
\caption{Feature importance of a Random Forest classifier that predicts if the $2_{nd}$ scholar would reciprocate or not to the $1_{st}$ scholar's inclusion his/her co-author panel.}
\label{tab:classifier_feature_importance}
 \end{table}

\section{Conclusion}
\label{sec:discussion}
Above all, in this work, we show that scientometrics and reciprocality can explain some of the unique socio-academic dynamics of co-author panel selection in GS profiles based on the self-identification theory. 

Starting with scientometrics, our results combine to suggest that, indeed, scholars tend to include co-authors with higher scientometrics, most notably \textit{lifetime} scientometrics (citation counts, H-index, and i10-index), over others in their co-authorship panel. This result is very much aligned with what one may expect based on the self-identification theory. Namely, the emphasis of one's association with prominent scholars is likely to reflect positively on that scholar as well \cite{ariel_6}. Furthermore, this tendency is significantly mitigated in our data with one's own scientometrics. In the context of self-identification theory, this result suggests that high-caliber scholars (i.e., those with high scientometrics) are less in need to \say{showcase} their associations with prominent scholars as they are, in fact, prominent themselves. In addition, it may be the case that these scholars' associations are already well-known or bare little effect on their status given their high scientometrics. Other factors, outside the scope of this study, may also partially explain this relation. For example, some more experienced scholars may be less technologically oriented and thus may put less effort into updating and maintaining their co-authorship panels over the years. Additionally, well-established scholars may wish to promote other aspects of their collaboration network such as the mentoring of young proteges, internationalism, and inter-disciplinarity, to name a few \cite{rosa_4}. We plan to investigate these and similar potential reasons in future work.

Turning to the issue of reciprocality, most inclusions in our data are reciprocated. While some differences were encountered based on discipline, other factors such as gender, academic age, and one's own scientometrics do not seem to significantly regulate this phenomenon. Taken jointly, these results suggest that while different disciplines may have slightly different norms, the concept of reciprocality in co-authorship panels is quite prevalent in practice in the academic community. One may consider this result to be intuitive -- since scholars tend to rely on each other's work, experience, and collaboration networks for a variety of professional purposes \cite{teddy_academia_1,rosa_5,rosa_6}, reciprocality natural emerges. However, given that scholars tend to associate themselves with high-performing scholars, reciprocality can be partially conflicting in cases where one of the scholars of the interaction is not what the other considers a prominent scholar. 

This study is not without limitations. First, our data is taken from a single platform (GS) which may present different characteristics than others due to its design and use patterns \cite{ariel_5}. Therefore, further investigation into additional different platforms is required. Second, our study primarily focused on the potential explanatory power of scientometrics and reciprocality in the selection of co-authorship panels. As a result, other potential factors such as research expertise, geographic proximity, and personal relationships, which are known to influence collaboration patterns \cite{ariel_4}, were not considered at all. We plan to consider a broader range of variables, as well as additional instruments such as surveys and interviews, to provide a more comprehensive understanding of co-author panel construction in future work. 

\section*{Declarations}
\subsection*{Funding}
This research did not receive any specific grant from funding agencies in the public, commercial, or not-for-profit sectors.

\subsection*{Conflicts of interest/Competing interests}
None.

\subsection*{Code and Data availability}
The code and data that have been used in this study are available upon reasonable request from the authors.

\subsection*{Author Contribution}
Ariel Alexi: Software, Data Curation, Formal analysis, Investigation, Writing - Review \& Editing. \\
Teddy Lazebnik: Resources, Software, Methodology, Project administration, Visualization, Writing - Original Draft, Writing - Review \& Editing, Supervision. \\ 
Ariel Rosenfeld: Conceptualization, Formal analysis, Validation, Writing - Original Draft, Writing - Review \& Editing, Supervision.\\
 
\bibliography{biblio}
\bibliographystyle{unsrt}

\end{document}